\documentclass[12pt]{article}
\usepackage[english]{babel}
\usepackage{amsmath,amsthm,amssymb,amsfonts}
\usepackage{color}

\newcommand{\sym}{\hbox{sym}}
\newcommand{\arctanh}{\hbox{arctanh}}

\begin{document}

\title{Thermodynamical analysis\\ and constitutive equations for a mixture\\ of viscous Korteweg fluids}
\author{M.~Gorgone, F.~Oliveri and P.~Rogolino\\
\ \\
{\footnotesize Department of Mathematical and Computer Sciences,}\\
{\footnotesize Physical Sciences and Earth Sciences, University of Messina}\\
{\footnotesize Viale F. Stagno d'Alcontres 31, 98166 Messina, Italy}\\
{\footnotesize mgorgone@unime.it; foliveri@unime.it; progolino@unime.it}
}

\date{Published in \textit{Phys. Fluids} \textbf{33}, 093102 (2021).}

\maketitle

\begin{abstract}
A complete thermodynamical analysis for a binary mixture of viscous Korteweg fluids with two velocities and two 
temperatures is developed. 
The constitutive functions are allowed to depend on the diffusion velocity and the specific internal 
energies of both constituents, together with their first gradients, on the symmetric part of the gradient of barycentric velocity,  as well as on the mass density of the mixture and the 
concentration of one of the constituents, together with their first and second gradients. Compatibility with 
entropy principle is analyzed by applying the extended Liu procedure, and a complete solution of the set of thermodynamical restrictions is recovered in three space dimensions.
Finally, the equilibrium configurations are investigated, and it is proved that no restrictions arise on the admissible phase boundaries. The theoretical results here provided may serve as a basis for experimental and/or numerical investigations, in particular for determining the surface levels of phase boundaries at equilibrium and making a comparison with experimental profiles.
\end{abstract}

\noindent
\textbf{Keywords.} Korteweg fluids; Binary mixtures with two velocities and two temperatures; Exploitation of second law of thermodynamics; Extended Liu procedure.

\section{Introduction}
\label{sec:intro}
The study of mixtures in the context of rational thermodynamics \cite{Mul1,GurVar,Bow,LiuMul} provides a 
framework for investigating several fundamental problems of continuum physics \cite{TRUE}.
For a  mixture made by $N\ge 2$ fluids, we have to determine the evolution of $5N$ 
fundamental fields, say the $N$ partial mass densities $\rho^{(a)}$ $(a=1,\ldots, N)$, the $N$  velocities $\mathbf{v}^{(a)}$, and the $N$ partial temperatures $\theta^{(a)}$ (or, equivalently, partial internal energies 
$\varepsilon^{(a)}$) of the constituents.
A standard procedure allows to define some quantities related to the mixture as a whole (mass density, barycentric 
velocity and internal energy of the mixture).

Both reacting and non-reacting mixtures can be considered; in this paper, we focus on non-reacting ones. Mixtures can be modeled at various degrees of detail (see, for instance, Refs.~\cite{BowGar,PPR2005,GouRug,BotDre,CGOP-2020}). The most classical models need to have as basic fields the partial mass densities of the constituents, the barycentric velocity and the temperature (or the internal energy) of the mixture
\cite{GurVar,LiuMul,GR-JNET}. From a thermodynamical viewpoint, one has to set properly the form of the local balances of energy and entropy. When non-local constitutive equations are considered, some authors added extra-terms in the energy equation (like the interstitial working, engendered by long-range interactions among the molecules  \cite{DunSer,Dun}), or in the entropy flux \cite{Mul}, in order to ensure the compatibility with second law of thermodynamics (see  Refs.~\cite{CimOliPac3,CimSelTri3} for an extensive discussion).

With a greater degree of detail we may consider the case where the constituents have different velocities, so that diffusive effects are taken into account;  nevertheless, the thermodynamical description of such mixtures remains at a classical level \cite{BotDre,FraPalRog,FraPalRog1}.

Finally, one may investigate the case where the constituents have their own temperature \cite{BowGar,GouRug}; such a situation may be relevant in plasma theories, where the different constituents of the plasma may experience different temperatures on time scales of the same order of magnitude of the transport process times.

In passing, we observe that mixtures whose description requires the introduction of internal state variables as 
additional fields  can be considered \cite{FraPalRog,FraPalRog1,OliPalRog}, even if this case will not be faced hereafter.

In this paper, we  consider a two-temperature and diffusive model of mixture made by two viscous Korteweg-like fluids, \emph{i.e.}, fluids whose state space includes also the second spatial derivatives of the mass density \cite{DunSer}.
The introduction of such a kind of fluids dates back to 1901 when the Dutch physicist Korteweg \cite{Kor}, in order 
to model fluid capillarity effects, proposed a constitutive equation for the Cauchy stress involving density gradients, namely
\begin{equation} 
\label{kort}
\mathbf{T}=\left(-p+\alpha_1|\nabla\rho|^2+\alpha_2\Delta\rho\right)\mathbf{I}+\alpha_3\nabla\rho\otimes\nabla\rho+\alpha_4\nabla\nabla\rho,
\end{equation}
where $\rho$ denotes the mass density, $p$ the pressure, $\mathbf{I}$ the identity matrix, and $\alpha_i$ ($i=1,\ldots,4$)  
suitable material functions depending on mass density and temperature
($[\alpha_1]=[\alpha_3]=Kg^{-1}m^7 s^{-2}$, $[\alpha_2]=[\alpha_4]=m^{4}\, s^{-2}$); moreover, $\Delta$ is the Laplacian operator, and the symbol $\otimes$ denotes tensorial product.
In the Korteweg's approach to capillarity,  interaction phenomena at the interfaces between liquid and vapor phases are described in terms of properties of an interfacial zone of finite thickness (diffuse interface) where density changes continuously \cite{Pan}. 
Diffuse interface models became of great interest in recent years in several fluid mechanics applications, such as phase transition phenomena \cite{Soucek-Heida-Malek,Swaney,Suli}.
Korteweg fluids have been theoretically investigated in a pioneeristic paper by Dunn and Serrin \cite{DunSer}, where the compatibility with the principles
of rational continuum thermodynamics \cite{TRUE} has been extensively studied,  by Cimmelli and coworkers  
\cite{CimOliPac3,CimSelTri3,CimSelTri1,CimOlTri,COP-CMT-2015,GOR-2020} through a generalized Liu procedure \cite{Liu,Cim1}, and by 
Heida and M\'alek \cite{HeiMal} following a different methodology. 
Korteweg-type models are also relevant in the modeling of granular materials \cite{Hutter-Raja}.

Remarkably, an explicit solution of the thermodynamical restrictions imposed by second law of thermodynamics in the general three-dimensional case of a viscous Korteweg fluid has been obtained in Ref.~\cite{GP-2021}. Korteweg fluids with the constitutive equation for Cauchy stress tensor given by relation (\ref{kort}) belong to a subclass of materials of grade 3. In general, a continuum material is said to be of grade $n$ if the constitutive quantities are allowed to depend 
on all gradients of the deformation of order equal to the integer $n$ \cite{True_Raja}. 

The plan of the paper is as follows. In Section~\ref{sec:balance}, starting from the governing equations for the two constituents, we derive the
evolution equations for the basic fields we choose to describe the mixture (mixture mass density, concentration of one constituent, barycentric and diffusion velocity, partial internal energies), together with the entropy inequality; the latter expresses, locally, the second law of thermodynamics. Then, in Section~\ref{sec:liu}, we sketch the generalized Liu procedure we apply for the exploitation of the entropy principle \cite{CimOlTri,Cim1}, and recover a set of conditions ensuring that second law of thermodynamics is satisfied for 
arbitrary thermodynamical processes. In Section~\ref{sec:3d}, we explicitly provide a solution to the restrictions on the constitutive equations. The lengthy computations necessary to derive and solve the compatibility restrictions placed by second law of thermodynamics are handled with the help of some symbolic routines written in the Computer Algebra System Reduce \cite{Reduce}. 
As a result, we are able to prove that second law of thermodynamics allows the dependence of the constitutive equations on all the gradients entering the state space, and so are compatible with a very general form of the Cauchy stress tensor which encompasses the constitutive equation of Korteweg fluids. 
The results achieved in this paper generalize those recently obtained in the one-dimensional case \cite{CGOP-2020}. Nevertheless, differently from the model considered in Ref.~\cite{CGOP-2020}, here we include  viscous terms in the Cauchy stress tensors. Moreover, the extension to the three-dimensional case carries out some more complications due to the constraints imposed by objectivity principle on the representation of scalar, vectorial and tensorial constitutive quantities; last but not the least, we have to be careful in the derivation of the thermodynamic constraints because of the symmetry of some tensorial quantities.
In Section~\ref{sec:serrin3D}, the problem of determining the phase boundaries at the equilibrium for the mixture is investigated; remarkably, it is proved that the recovered constitutive quantities are such that we do not have any limitation to the phase boundaries
at the equilibrium. This problem does not occur in the one-dimensional situation considered in Ref.~\cite{CGOP-2020} where the equations for equilibrium are not overdetermined.
Finally, Section~\ref{sec:final} our concluding remarks as well as possible 
developments of the present theory.

\section{Field equations and entropy inequality}
\label{sec:balance}
Let us consider a non-reacting binary mixture of Korteweg fluids. 
In the absence of external forces and  heat 
sources, and neglecting momentum and energy exchanges between the components, the field equations for each constituent are
\begin{equation}
\begin{aligned}
&\frac{\partial \rho^{(a)}}{\partial t}+\nabla\cdot \left(\rho^{(a)}\mathbf{v}^{(a)}\right)=0,\\
&\rho^{(a)}\left(\frac{\partial\mathbf{v}^{(a)}}{\partial t}+(\mathbf{v}^{(a)}\cdot\nabla)\mathbf{v}^{(a)}\right)-\nabla\cdot \mathbf{T}^{(a)}=\mathbf{0},\\
&\rho^{(a)}\left(\frac{\partial \varepsilon^{(a)}}{\partial t}+\mathbf{v}^{(a)}\cdot\nabla\varepsilon^{(a)}\right)-\mathbf{T}^{(a)}\cdot \nabla\mathbf{v}^{(a)}+\nabla\cdot \mathbf{q}^{(a)}=0,
\end{aligned}
\end{equation}
where the superscript ${}^{(a)}$ ($a=1,2$) labels the two constituents, $\rho^{(a)}(t,\mathbf{x})$ denotes the mass density,
$\mathbf{v}^{(a)}(t,\mathbf{x})$ the velocity, $\varepsilon^{(a)}(t,\mathbf{x})$ the internal energy per unit mass, 
$\mathbf{T}^{(a)}$ the symmetric Cauchy stress tensor, and $\mathbf{q}^{(a)}$ the heat flux; here and in the sequel, $\mathbf{A}\cdot\mathbf{B}$ denotes the full contraction of vectors and tensors, \emph{i.e.}, $\mathbf{A}\cdot\mathbf{B}=\hbox{tr}(\mathbf{A}^T\mathbf{B})$, where the superscript ${}^T$ stands for transposition.

It is worth observing that, since the velocities of the components are different, the convective time derivatives for the two fluids are not the same. 
Let us introduce the mass density $\rho$ of the whole mixture together with the 
concentration $c$ of the first constituent  \cite{GurVar,LiuMul},  as well as the barycentric velocity $\mathbf{v}$ of the whole mixture, and the diffusion velocity $\mathbf{w}$ of the first constituent, namely 
\begin{equation}
\begin{aligned}
&\rho=\rho^{(1)}+\rho^{(2)}, \qquad &&c=\frac{\rho^{(1)}}{\rho},\qquad  0<c<1,\\
&\mathbf{v}= \frac{\rho^{(1)} \mathbf{v}^{(1)}+\rho^{(2)} \mathbf{v}^{(2)}}{\rho}, \qquad 
&&\mathbf{w}=\mathbf{v}^{(1)}-\mathbf{v}.
\end{aligned}
\end{equation}
As basic fields to describe the binary mixture we take $\rho$, $c$, $\mathbf{v}$, $\mathbf{w}$, $\varepsilon^{(1)}$ and $\varepsilon^{(2)}$,
whereupon we are led to consider the following evolution equations for the binary mixture: 
\begin{equation}
\label{modelequations}
\begin{aligned}
&\mathcal{E}^{(1)}\equiv\frac{\partial \rho}{\partial t}+\nabla\cdot (\rho\mathbf{v})=0,\\
&\mathcal{E}^{(2)}\equiv\rho\left(\frac{\partial c}{\partial t}+\mathbf{v}\cdot \nabla \mathbf{c}\right)+\nabla\cdot(\rho c \mathbf{w})=0,\\
&\mathcal{E}^{(3)}\equiv\rho\left(\frac{\partial\mathbf{v}}{\partial t}+(\mathbf{v}\cdot\nabla)\mathbf{v}\right)-\nabla\cdot\left( \mathbf{T}^{(1)} +\mathbf{T}^{(2)}-\frac{\rho c}{1-c}\mathbf{w}\otimes\mathbf{w}\right)=\mathbf{0},\\
&\mathcal{E}^{(4)}\equiv\rho c\left(\frac{\partial\mathbf{w}}{\partial t}+\left((\mathbf{v}+\mathbf{w})\cdot\nabla\right)\mathbf{w}+(\mathbf{w}\cdot\nabla)\mathbf{v}\right)\\
&\qquad-\nabla\cdot\mathbf{T}^{(1)}+c\nabla\cdot\left(\mathbf{T}^{(1)}+\mathbf{T}^{(2)}-\frac{\rho c}{1-c} \mathbf{w}\otimes\mathbf{w}\right)=\mathbf{0},\\
&\mathcal{E}^{(5)}\equiv\rho c\left(\frac{\partial \varepsilon^{(1)}}{\partial t}+(\mathbf{v}+\mathbf{w})\cdot \nabla \varepsilon^{(1)}\right)-\mathbf{T}^{(1)}\cdot\nabla(\mathbf{v}+\mathbf{w})
+\nabla\cdot\mathbf{q}^{(1)}=0,\\
&\mathcal{E}^{(6)}\equiv\rho (1-c)\left(\frac{\partial \varepsilon^{(2)}}{\partial t}+
\left(\mathbf{v}-\frac{c}{1-c}\mathbf{w}\right)\cdot\nabla\varepsilon^{(2)}\right)\\
&\qquad-\mathbf{T}^{(2)}\cdot\nabla\left(\mathbf{v}-\frac{c}{1-c}\mathbf{w}\right)+\nabla\cdot \mathbf{q}^{(2)}=0.
\end{aligned}
\end{equation}
The choice of the barycentric velocity $\mathbf{v}$ and the diffusion velocity $\mathbf{w}$ in place of the velocities $\mathbf{v}^{(1)}$ and $\mathbf{v}^{(2)}$ of the two fluids is motivated by technical reasons due to the integration of thermodynamic constraints: in fact, the state space cannot include separately $\mathbf{v}^{(1)}$ and $\mathbf{v}^{(2)}$ because only their difference is an objective quantity; on the contrary, $\mathbf{w}$ can belong to the state space. These choices do not determine any loss of generality in the thermodynamical analysis; moreover, the use of $\rho$ and $c$ in place of $\rho^{(1)}$ and $\rho^{(2)}$ does not affect the results as well. Of course, the constitutive quantities we obtain can be rewritten in terms of the fields related to the constituents (this is explicitly done when considering the equilibrium conditions).

As far as the momentum equation $\mathcal{E}^{(3)}$ for the whole mixture is concerned, we observe that we may define the Cauchy stress tensor for the mixture, say
\begin{equation*}
\label{T_total}
\mathbf{T}=\mathbf{T}^{(1)} +\mathbf{T}^{(2)}-\frac{\rho c}{1-c}\mathbf{w}\otimes\mathbf{w}.
\end{equation*}
Moreover, the energy equation for the whole mixture,
\begin{equation*}
\rho\left(\frac{\partial \varepsilon}{\partial t}+\mathbf{v}\cdot\nabla\varepsilon\right)-\mathbf{T}\cdot \nabla\mathbf{v}+\nabla\cdot\mathbf{q}=0,
\end{equation*}
can be obtained by combining the energy equations for the two components, along with
$\rho\varepsilon=\rho^{(1)}\varepsilon^{(1)}+\rho^{(2)}\varepsilon^{(2)}$, and the total heat flux given by
\begin{equation*}
\label{q_total}
\mathbf{q}=\mathbf{q}^{(1)}+\mathbf{q}^{(2)}-\left(\mathbf{T}^{(1)}+\frac{c}{1-c}\mathbf{T}^{(2)}-\rho c\left(\varepsilon^{(1)}-\varepsilon^{(2)}+\frac{1-2c}{2(1-c)^2}|\mathbf{w}|^2\right)\right)\mathbf{w}.
\end{equation*}
It is easily recognized that in the case of a non-diffusive mixture, \emph{i.e.}, $\mathbf{w}=\mathbf{0}$, the Cauchy stress tensor and the heat flux of the whole mixture are  just the sum of their partial contributions; however, in this case, taking into account equation~(\ref{modelequations})$_4$, the two partial Cauchy stress tensors must satisfy the condition
\begin{equation*}
(1-c)\nabla\cdot\mathbf{T}^{(1)}-c\nabla\cdot\mathbf{T}^{(2)}=\mathbf{0}.
\end{equation*}
In the limiting cases $c\rightarrow 1$ (the mixture reduces to the first fluid component), we have 
\[
\mathbf{v}\rightarrow \mathbf{v}^{(1)}\quad \hbox{and}\quad \mathbf{w}\rightarrow \mathbf{0}, 
\]
whereas for $c\rightarrow 0$ (the mixture reduces to the second fluid component), it is
\[
\mathbf{v}\rightarrow \mathbf{v}^{(2)}\quad \hbox{and}\quad \mathbf{w}\rightarrow -\mathbf{v}.
\]

Finally, we consider the entropy inequality for the whole mixture that in local form reads as
\begin{equation}
\label{entropyinequality}
\rho \left(\frac{\partial s}{\partial t}+\mathbf{v}\cdot\nabla s\right)+\nabla\cdot \mathbf{J}\ge 0,
\end{equation}
where $s$ is the specific entropy, and $\mathbf{J}$ is the entropy flux.

The field equations (\ref{modelequations}) and the entropy inequality (\ref{entropyinequality}), once the variables entering the state space have been assigned, must be 
supplemented by the constitutive equations for partial Cauchy stress tensors, partial heat fluxes, specific entropy and 
entropy flux.
In view of equation~(\ref{kort}),  let us assume the state space 
to be spanned by
\begin{equation}
\label{statespace}
\mathcal{Z}=\left\{\rho,c,\mathbf{w},\varepsilon^{(1)},\varepsilon^{(2)},\nabla\rho,
\nabla c,\mathbf{L},\nabla\mathbf{w},\nabla\varepsilon^{(1)},
\nabla\varepsilon^{(2)},\nabla\nabla\rho,\nabla\nabla c\right\},
\end{equation}
where $\mathbf{L}=\sym(\nabla\mathbf{v})$,
\emph{i.e.}, we consider a second order non-local theory.

The entropy principle imposes the inequality (\ref{entropyinequality}) be satisfied for arbitrary 
thermodynamical processes \cite{CimJouRugVan,JCL}. To find a set of conditions which are at least sufficient for the 
fulfillment of such a requirement, we apply an extended Liu procedure 
\cite{CimOlTri,Cim1}, incorporating new restrictions consistent with higher order non-local constitutive theories.
According to this procedure, in the entropy inequality, we have to use as constraints the field equations, and their gradient extensions too, up to the order of the gradients entering the state space, by means of suitable Lagrange multipliers.

There is a simple \emph{rationale} for using the gradients of the governing equations
as additional constraints in the entropy inequality  when dealing with non-local constitutive equations. 
Thermodynamical processes are solutions of the field equations, and, if these solutions are smooth enough, 
are trivially solutions of their differential consequences  (see also Ref.~\cite{RogCim}). 
Since the entropy inequality (\ref{entropyinequality}) has to be satisfied in arbitrary smooth processes,  then, from a mathematical viewpoint, we have to use also the differential consequences of the equations governing those 
processes as constraints for such an inequality. On the contrary, if we limit ourselves to consider as constraints 
only the field equations, we are led straightforwardly to a specific entropy and Lagrange multipliers which are 
independent of the gradients of the variables entering the state space \cite{CGOP-2020,GP-2021}. As a direct consequence, a Cauchy stress tensor depending on the 
gradients of mass density should be incompatible with second law of thermodynamics. 

It is worth of being remarked that we assume the energy equations in the classical form 
(\emph{i.e.}, no interstitial working \cite{DunSer} is introduced)
and do not postulate the functional form of entropy flux; remarkably, an entropy flux given by the classical term and some additional contributions will arise in a natural way from the algorithmic application of the procedure itself.

\section{Extended Liu procedure and thermodynamical restrictions}
\label{sec:liu}
In order to exploit second law of thermodynamics, we take into account the constraints 
imposed on entropy inequality by the field equations and their gradient extensions; this task is accomplished by introducing the Lagrange multipliers $\lambda^{(1)}$,  $\lambda^{(2)}$, $\boldsymbol\lambda^{(3)}\equiv\{\lambda^{(3)}_i, i=1,2,3\}$, $\boldsymbol\lambda^{(4)}\equiv \{\lambda^{(4)}_i, i=1,2,3\}$, $\lambda^{(5)}$, $\lambda^{(6)}$, $\boldsymbol\Lambda^{(1)}\equiv\{\Lambda^{(1)}_i, i=1,2,3\}$, $\boldsymbol\Lambda^{(2)}\equiv 
\{\Lambda^{(2)}_i, i=1,2,3\}$,
$\boldsymbol\Lambda^{(3)}\equiv\{\Lambda^{(3)}_{ij}, i,j=1,2,3\}$, $\boldsymbol\Lambda^{(4)}\equiv\{\Lambda^{(4)}_{ij}, i,j=1,2,3\}$, $\boldsymbol\Lambda^{(5)}\equiv\{\Lambda^{(5)}_{i}, i=1,2,3\}$, $\boldsymbol\Lambda^{(6)}\equiv\{ \Lambda^{(6)}_{i}, i=1,2,3\}$, $\boldsymbol\Theta^{(1)}\equiv\{
\Theta^{(1)}_{ij}, i,j=1,2,3\}$ and $\boldsymbol\Theta^{(2)}\equiv\{
\Theta^{(2)}_{ij}, i,j=1,2,3\}$, depending on the state space variables.
Thus, the entropy inequality writes
\begin{equation}
\label{entropyconstrained}
\begin{aligned}
&\rho \left(\frac{\partial s}{\partial t}+\mathbf{v}\cdot \nabla s\right)+\nabla\cdot \mathbf{J} \\
&\quad- \lambda^{(1)}\mathcal{E}^{(1)}- \lambda^{(2)}\mathcal{E}^{(2)}
- \boldsymbol\lambda^{(3)}\cdot\mathcal{E}^{(3)}- \boldsymbol\lambda^{(4)}\cdot\mathcal{E}^{(4)}
-\lambda^{(5)}\mathcal{E}^{(5)}- \lambda^{(6)} \mathcal{E}^{(6)}\\
&\quad-\boldsymbol\Lambda^{(1)}\cdot\nabla\mathcal{E}^{(1)}
-\boldsymbol\Lambda^{(2)}\cdot\nabla\mathcal{E}^{(2)} 
-\boldsymbol\Lambda^{(3)}\cdot\nabla\mathcal{E}^{(3)}
-\boldsymbol\Lambda^{(4)}\cdot\nabla\mathcal{E}^{(4)}\\
&\quad-\boldsymbol\Lambda^{(5)}\cdot\nabla\mathcal{E}^{(5)}
-\boldsymbol\Lambda^{(6)}\cdot\nabla\mathcal{E}^{(6)}-\boldsymbol\Theta^{(1)}\cdot\nabla\nabla\mathcal{E}^{(1)}
-\boldsymbol\Theta^{(2)}\cdot\nabla\nabla \mathcal{E}^{(2)}\geq0.
\end{aligned}
\end{equation}

The inequality (\ref{entropyconstrained}) needs to be expanded by means of the chain rule; nevertheless, since the computations though straightforward are tremendously long (in fact, in the present case the expanded entropy inequality involves 1074395 terms!), we perform the task by using some routines written in the Computer Algebra System Reduce \cite{Reduce}, and we omit to report the full form here. The main advantages of using these symbolic routines are that we are able to extract the coefficients of a multivariate polynomial in some derivatives of the field variables, and then solve, with the help of the Crack package \cite{Wolf}, the set of differential and algebraic conditions for 
the unknown constitutive functions. 

In the expanded version of (\ref{entropyconstrained}), we can distinguish the \emph{highest derivatives} and the \emph{higher derivatives}  \cite{CimSelTri1}. 
The formers are both the time derivatives of the field variables and of the
elements of the state space, which cannot be expressed through the governing equations as functions of the 
thermodynamical variables, and the spatial derivatives whose order is the highest one. On the contrary, the higher derivatives are the spatial derivatives whose order is not maximal but higher than that of the gradients entering the 
state space. In the following, for the generic field $u(t,\mathbf{x})$, let us use the following compact notation for derivatives:
$\displaystyle u_{,t}=\frac{\partial u}{\partial t}$, $\displaystyle u_{,i}=\frac{\partial u}{\partial x_i}$, 
$\displaystyle u_{,it}=\frac{\partial^2 u}{\partial x_i\partial t}$, $\displaystyle u_{,ij}=\frac{\partial^2 u}{\partial x_i\partial x_j}$, \ldots.
As a consequence of  the choice of the state space (\ref{statespace}), the components of the highest derivatives are 
\begin{equation}
\begin{aligned}
\mathbf{X}=&\left\{\rho_{,t},\;c_{,t},\;v_{i,t},\;w_{i,t},\;\varepsilon^{(1)}_{,t},\;\varepsilon^{(2)}_{,t},\;\rho_{,kt},\;c_{,kt},\;v_{i,kt},\;w_{i,kt},\;
\varepsilon^{(1)}_{,kt},\;\varepsilon^{(2)}_{,kt},\right. \\
&\;\;\,\left.\rho_{,k\ell t},\;c_{,k\ell t},\;v_{i,jk\ell},\;
w_{i,jk\ell},\;\varepsilon^{(1)}_{,jk\ell},\;\varepsilon^{(2)}_{,jk\ell},\;\rho_{,jk\ell m},\;c_{,jk\ell m}\right\},
\end{aligned}
\end{equation}
whereas the components of the higher ones are
\begin{equation}
\mathbf{Y}=\left\{v_{i,jk},\;w_{i,jk},\varepsilon^{(1)}_{,jk},\;\varepsilon^{(2)}_{,jk},\;\rho_{,jk\ell},\;c_{,jk\ell}\right\};
\end{equation}
let $n_1$ denote the number of the components of the highest derivatives, and $n_2$ the number of the components of higher derivatives.
Using these positions, the entropy inequality (\ref{entropyconstrained}) can be recasted in the compact form
\begin{equation}
\label{entropycompatta}
\mathbf{A}\cdot\mathbf{X}+\mathbf{Y}^{T}\mathbf{B}\mathbf{Y}+\mathbf{C}\cdot\mathbf{Y}+D\ge 0,
\end{equation}
where $\mathbf{B}$ is a symmetric matrix of order $n_2$, $\mathbf{A}$ is a vector with $n_1$ components, $\mathbf{C}$ is a vector with $n_2$ components, and $D$ is a scalar; $\mathbf{A}$, $\mathbf{B}$, $\mathbf{C}$ and $D$ depend at most on the field variables and the gradients entering the state space.   
Therefore, the left hand side of the inequality (\ref{entropycompatta}) is a polynomial which is linear in the highest derivatives and quadratic in the higher ones. Because of the constraints we imposed,  the highest and higher derivatives may assume arbitrary values, whereupon, in order to fulfill the second law of thermodynamics, inequality~(\ref{entropycompatta}) must hold for arbitrary $\mathbf{X}$ and $\mathbf{Y}$. Therefore, the coefficients of linear terms in the highest and higher derivatives must vanish, otherwise, the entropy inequality could be easily violated. Since in principle nothing prevents the possibility of a thermodynamical process where $D=0$,  in order the inequality (\ref{entropycompatta}) be satisfied for every thermodynamical process, 
the conditions
\begin{equation}
\mathbf{A}=\mathbf{0}, \qquad \mathbf{C}=\mathbf{0},\qquad D\ge 0,
\end{equation} 
together with the requirement that $\mathbf{B}$ is a positive semidefinite matrix, are  sufficient for the fulfillment of the entropy inequality \cite{CimOlTri}.
These sufficient conditions provide a set of constraints on the constitutive equations.
 
From $\mathbf{A}=\mathbf{0}$, we obtain the expressions for the components of Lagrange multipliers,
\begin{equation}
\label{lagrange}
\begin{aligned}
&\lambda^{(1)}=\rho\frac{\partial s}{\partial\rho},\qquad\lambda^{(2)}=\frac{\partial s}{\partial c}-\frac{\rho_{,k}}{\rho}\left(\frac{\partial s}{\partial c_{,k}}-2\frac{\rho_{,i}}{\rho}\frac{\partial s}{\partial c_{,ik}}\right)-\frac{\rho_{,ik}}{\rho}\frac{\partial s}{\partial c_{,ik}},\\
&\lambda^{(3)}_i=-\frac{\rho_{,k}}{\rho}\frac{\partial s}{\partial v_{i,k}},\qquad\lambda^{(4)}_i=\frac{1}{c}\frac{\partial s}{\partial w_i}-\frac{\rho c_{,k}+c\rho_{,k}}{\rho c^2}\frac{\partial s}{\partial w_{i,k}}, \\
&\lambda^{(5)}=\frac{1}{c}\frac{\partial s}{\partial \varepsilon^{(1)}}-\frac{\rho c_{,k}+c\rho_{,k}}{\rho c^2}\frac{\partial s}{\partial \varepsilon^{(1)}_{,k}}, \\
&\lambda^{(6)}=\frac{1}{1-c}\frac{\partial s}{\partial \varepsilon^{(2)}}-\frac{(1-c)\rho_{,k}-\rho c_{,k}}{\rho(1-c)^2}\frac{\partial s}{\partial \varepsilon^{(2)}_{,k}}, \\
&\Lambda^{(1)}_k=\rho\frac{\partial s}{\partial\rho_{,k}},\qquad\Lambda^{(2)}_k=\frac{\partial s}{\partial c_{,k}}-2\frac{\rho_{,i}}{\rho}\frac{\partial s}{\partial c_{,ik}}, \\
&\Lambda^{(3)}_{ik}=\frac{\partial s}{\partial v_{i,k}},\qquad\Lambda^{(4)}_{ik}=\frac{1}{c}\frac{\partial s}{\partial w_{i,k}}, \qquad
\Lambda^{(5)}_{k}=\frac{1}{c}\frac{\partial s}{\partial \varepsilon^{(1)}_{,k}},\\
&\Lambda^{(6)}_{k}=\frac{1}{1-c}\frac{\partial s}{\partial \varepsilon^{(2)}_{,k}}, \qquad
\Theta^{(1)}_{ik}=\rho\frac{\partial s}{\partial\rho_{,ik}},\qquad\Theta^{(2)}_{ik}=\frac{\partial s}{\partial c_{,ik}},
\end{aligned}
\end{equation}
as well as the following restrictions involving the specific entropy, partial Cauchy stress tensors and heat fluxes:
\begin{equation}
\label{restricthighest1}
\begin{aligned}
&\left\langle\frac{\partial s}{\partial v_{i,k}}\left(\frac{\partial T^{(1)}_{ij}}{\partial v_{\ell,m}}+\frac{\partial T^{(2)}_{ij}}{\partial v_{\ell,m}}\right)+\frac{1}{c}\frac{\partial s}{\partial w_{i,k}}\left((1-c)\frac{\partial T^{(1)}_{ij}}{\partial v_{\ell,m}}-c\frac{\partial T^{(2)}_{ij}}{\partial v_{\ell,m}}\right)\right. \\
&\qquad
-\left.\frac{1}{c}\frac{\partial s}{\partial \varepsilon^{(1)}_{,k}}\frac{\partial q^{(1)}_{j}}{\partial v_{\ell,m}}-\frac{1}{1-c}\frac{\partial s}{\partial \varepsilon^{(2)}_{,k}}\frac{\partial q^{(2)}_{j}}{\partial 
v_{\ell,m}}-\rho^2\frac{\partial s}{\partial\rho_{,km}}\delta_{j\ell}\right\rangle_{(jk\ell m)}=0, 
\end{aligned}
\end{equation}
\begin{equation}
\label{restricthighest2}
\begin{aligned}
&\left\langle\frac{\partial s}{\partial v_{i,k}}\left(\frac{\partial T^{(1)}_{ij}}{\partial w_{\ell,m}}+\frac{\partial T^{(2)}_{ij}}{\partial w_{\ell,m}}\right)+\frac{1}{c}\frac{\partial s}{\partial w_{i,k}}\left((1-c)\frac{\partial T^{(1)}_{ij}}{\partial w_{\ell,m}}-c\frac{\partial T^{(2)}_{ij}}{\partial w_{\ell,m}}\right)\right.\\
&\qquad
-\left.\frac{1}{c}\frac{\partial s}{\partial \varepsilon^{(1)}_{,k}}\frac{\partial q^{(1)}_{j}}{\partial w_{\ell,m}}-\frac{1}{1-c}\frac{\partial s}{\partial \varepsilon^{(2)}_{,k}}\frac{\partial q^{(2)}_{j}}{\partial 
w_{\ell,m}}-\rho c\frac{\partial s}{\partial c_{,km}}\delta_{j\ell}\right\rangle_{(jk\ell m)}=0, 
\end{aligned}
\end{equation}
\begin{equation}
\label{restricthighest3}
\begin{aligned}&\left\langle\frac{\partial s}{\partial v_{i,k}}\left(\frac{\partial T^{(1)}_{ij}}{\partial \varepsilon^{(1)}_{,\ell}}+\frac{\partial T^{(2)}_{ij}}{\partial \varepsilon^{(1)}_{,\ell}}\right)+\frac{1}{c}\frac{\partial s}{\partial w_{i,k}}\left((1-c)\frac{\partial T^{(1)}_{ij}}{\partial \varepsilon^{(1)}_{,\ell}}-c\frac{\partial T^{(2)}_{ij}}{\partial \varepsilon^{(1)}_{,\ell}}\right)\right. \\
&\qquad
-\left.\frac{1}{c}\frac{\partial s}{\partial \varepsilon^{(1)}_{,k}}\frac{\partial q^{(1)}_{j}}{\partial \varepsilon^{(1)}_{,\ell}}-\frac{1}{1-c}\frac{\partial s}{\partial \varepsilon^{(2)}_{,k}}\frac{\partial q^{(2)}_{j}}{\partial 
\varepsilon^{(1)}_{,\ell}}\right\rangle_{(jk\ell)}=0, 
\end{aligned}
\end{equation}
\begin{equation}
\label{restricthighest4}
\begin{aligned}&\left\langle\frac{\partial s}{\partial v_{i,k}}\left(\frac{\partial T^{(1)}_{ij}}{\partial \varepsilon^{(2)}_{,\ell}}+\frac{\partial T^{(2)}_{ij}}{\partial \varepsilon^{(2)}_{,\ell}}\right)+\frac{1}{c}\frac{\partial s}{\partial w_{i,k}}\left((1-c)\frac{\partial T^{(1)}_{ij}}{\partial \varepsilon^{(2)}_{,\ell}}-c\frac{\partial T^{(2)}_{ij}}{\partial \varepsilon^{(2)}_{,\ell}}\right)\right. \\
&\qquad
-\left.\frac{1}{c}\frac{\partial s}{\partial \varepsilon^{(1)}_{,k}}\frac{\partial q^{(1)}_{j}}{\partial \varepsilon^{(2)}_{,\ell}}-\frac{1}{1-c}\frac{\partial s}{\partial \varepsilon^{(2)}_{,k}}\frac{\partial q^{(2)}_{j}}{\partial 
\varepsilon^{(2)}_{,\ell}}\right\rangle_{(jk\ell)}=0, 
\end{aligned}
\end{equation}
\begin{equation}
\label{restricthighest5}
\begin{aligned}&\left\langle\frac{\partial s}{\partial v_{i,k}}\left(\frac{\partial T^{(1)}_{ij}}{\partial \rho_{,\ell m}}+\frac{\partial T^{(2)}_{ij}}{\partial \rho_{,\ell m}}\right)+\frac{1}{c}\frac{\partial s}{\partial w_{i,k}}\left((1-c)\frac{\partial T^{(1)}_{ij}}{\partial \rho_{,\ell m}}-c\frac{\partial T^{(2)}_{ij}}{\partial \rho_{,\ell m}}\right)\right. \\
&\qquad
-\left.\frac{1}{c}\frac{\partial s}{\partial \varepsilon^{(1)}_{,k}}\frac{\partial q^{(1)}_{j}}{\partial \rho_{,\ell m}}-\frac{1}{1-c}\frac{\partial s}{\partial \varepsilon^{(2)}_{,k}}\frac{\partial q^{(2)}_{j}}{\partial 
\rho_{,\ell m}}\right\rangle_{(jk\ell m)}=0, 
\end{aligned}
\end{equation}
\begin{equation}
\label{restricthighest6}
\begin{aligned}&\left\langle\frac{\partial s}{\partial v_{i,k}}\left(\frac{\partial T^{(1)}_{ij}}{\partial c_{,\ell m}}+\frac{\partial T^{(2)}_{ij}}{\partial c_{,\ell m}}\right)+\frac{1}{c}\frac{\partial s}{\partial w_{i,k}}\left((1-c)\frac{\partial T^{(1)}_{ij}}{\partial c_{,\ell m}}-c\frac{\partial T^{(2)}_{ij}}{\partial c_{,\ell m}}\right)\right. \\
&\qquad
-\left.\frac{1}{c}\frac{\partial s}{\partial \varepsilon^{(1)}_{,k}}\frac{\partial q^{(1)}_{j}}{\partial c_{,\ell m}}-\frac{1}{1-c}\frac{\partial s}{\partial \varepsilon^{(2)}_{,k}}\frac{\partial q^{(2)}_{j}}{\partial 
c_{,\ell m}}\right\rangle_{(jk\ell m)}=0,
\end{aligned}
\end{equation}
where the symbol $\langle\mathcal{F}\rangle_{(i_1\ldots i_r)}$ denotes the symmetric part of the tensor function $\mathcal{F}$ with respect to the indices $i_1\ldots i_r$. 

It is easily ascertained by direct inspection of relations (\ref{lagrange}) that the Lagrange multipliers and, hence, the specific entropy, are allowed to depend on the gradients of the unknown variables, and the same holds true for the partial Cauchy stress tensors. Moreover, as far as the relations (\ref{restricthighest1})--(\ref{restricthighest6}) are concerned, once we assign a representation for partial Cauchy stress tensors and heat fluxes, the dependence of specific entropy on the gradients entering the state space is somehow constrained. 

The further thermodynamical restrictions, even if their computation is straightforward, are omitted since their expression is rather long.
Furthermore, analyzing the conditions coming from $\mathbf{C}=\mathbf{0}$, it follows that the entropy flux is no longer given by the constitutive equation postulated in rational thermodynamics and an entropy
extra-flux is obtained. Indeed, in principle, this result could also be recovered in the classical Liu procedure, since the  Liu theorem  states that the Lagrange multipliers are defined on the whole state space \cite{Liu}. However, in this case, such a dependence does not  imply an analogous dependence of the partial Cauchy stress tensors on the gradients of the field variables, and the classical features of Korteweg-type fluids are lost. 
As a final comment, we observe that the thermodynamic restrictions are obtained by the exploitation of a rigorous mathematical procedure; due to their length, we do not try to attribute to them a detailed physical meaning. Nevertheless, the effectiveness of the procedure can be judged on the basis of the physical admissibility of the results it produces. 

In fact, in the next section, we give a solution to all the restrictions by determining explicit constitutive equations for the partial Cauchy stress tensors, the partial heat fluxes, the specific entropy and the 
entropy flux, so proving that the thermodynamical constraints recovered by applying the extended Liu procedure can be effectively solved, providing a complete and physically meaningful solution.

\section{Solution of thermodynamical constraints}
\label{sec:3d}
All the thermodynamical restrictions derived in Section~\ref{sec:liu} are still too general for being useful in concrete applications; therefore, a further simplification is necessary according to specific models we want to manage. Thus, in order to proceed with the exploitation of the entropy inequality for a binary mixture of viscous Korteweg fluids, let us assume the following constitutive equations for the partial Cauchy stress tensors and heat fluxes:
\begin{equation}
\label{consteq}
\begin{aligned}
\mathbf{T}^{(1)}&=\left(\tau^{(1)}_0+\tau^{(1)}_1 |\nabla\rho|^2+\tau^{(1)}_2 \nabla\rho\cdot\nabla c+\tau^{(1)}_3 |\nabla c|^2+\tau^{(1)}_4 \Delta\rho\right. \\
&+\left.\tau^{(1)}_5 \Delta c+\tau^{(1)}_6\nabla\cdot(\mathbf{v}+\mathbf{w})\right)\mathbf{I} 
\\
&+\tau^{(1)}_7\nabla\rho\otimes\nabla\rho+\tau^{(1)}_8\sym\left(\nabla\rho\otimes\nabla c\right)
+\tau^{(1)}_9\nabla c\otimes\nabla c \\
&+\tau^{(1)}_{10}\nabla\nabla\rho+\tau^{(1)}_{11}\nabla\nabla c+\tau^{(1)}_{12}\sym\left(\nabla\left(\mathbf{v}+\mathbf{w}\right)\right), \\
\mathbf{T}^{(2)}&=\left(\tau^{(2)}_0+\tau^{(2)}_1 |\nabla\rho|^2+\tau^{(2)}_2 \nabla\rho\cdot\nabla c+\tau^{(2)}_3 |\nabla c|^2+\tau^{(2)}_4 \Delta\rho\right. \\
&+\left.\tau^{(2)}_5 \Delta c+\tau^{(2)}_6\nabla\cdot\left(\mathbf{v}-\frac{c}{1-c}\mathbf{w}\right)\right)\mathbf{I} \\
&+\tau^{(2)}_7\nabla\rho\otimes\nabla\rho+\tau^{(2)}_8\sym\left(\nabla\rho\otimes\nabla c\right)+\tau^{(2)}_9\nabla c\otimes\nabla c \\
&+\tau^{(2)}_{10}\nabla\nabla\rho+\tau^{(2)}_{11}\nabla\nabla c+\tau^{(2)}_{12}\sym\left(\nabla\left(\mathbf{v}-\frac{c}{1-c}\mathbf{w}\right)\right), \\
\mathbf{q}^{(1)}&=q^{(1)}_1\nabla\varepsilon^{(1)}+q^{(1)}_2\nabla\varepsilon^{(2)},\\
\mathbf{q}^{(2)}&=q^{(2)}_1\nabla\varepsilon^{(1)}+q^{(2)}_2\nabla\varepsilon^{(2)},
\end{aligned}
\end{equation}
where $q^{(a)}_j$ ($j=1,2$) and $\tau^{(a)}_k$ ($k=0,\ldots,12$), with $a=1,2$, are suitable scalar material functions depending at most on $(\rho,c,|\mathbf{w}|,\varepsilon^{(1)},\varepsilon^{(2)})$.

Moreover,  let us expand the specific entropy $s$
around the homogeneous equilibrium state (where all gradients vanish), at the first order on the gradients of mass density of the whole mixture and concentration of the first constituent, \emph{i.e.},
\begin{equation}
\label{specificentropy}
s=\hat{s}_0+\hat{s}_1|\nabla\rho|^2+\hat{s}_2\nabla\rho\cdot\nabla c+\hat{s}_3|\nabla c|^2,
\end{equation}
where $\hat{s}_i\equiv \hat{s}_i(\rho,c,|\mathbf{w}|,\varepsilon^{(1)},\varepsilon^{(2)})$ $(i=0,\ldots,3)$ are some 
scalar functions of the indicated arguments. We are aware that expression (\ref{specificentropy}) is not the most general 
representation of
the entropy density as an isotropic scalar function, but it is enough for our purposes. The presence  of terms involving the gradients of $\rho$ and $c$ in the expression of the specific entropy is crucial for the compatibility of non-local Cauchy stress tensors with the second law of thermodynamics\cite{CGOP-2020}.

On the basis of the assumptions on constitutive relations (\ref{consteq}) and (\ref{specificentropy}), the thermodynamical constraints provide
a large set of partial differential equations that we manage
using some routines written in the Computer Algebra System
Reduce \cite{Reduce}; then, we are able to solve the restrictions (\ref{restricthighest1})--(\ref{restricthighest6}) so obtaining the form of the coefficients entering the specific entropy:
\begin{equation}
\label{solentropy}
\begin{aligned}
\hat{s}_0&=\frac{1}{\rho}\left(\rho c s_{01}+\rho(1-c)s_{02}+\phi_1+\phi_2\right), \\
\hat{s}_1&=\frac{\partial^3 s_1}{\partial\rho^3},\qquad\hat{s}_2=\rho\frac{\partial^3 s_2}{\partial\rho^3},\qquad
\hat{s}_3=\frac{\partial}{\partial\rho}\left(\rho^2\frac{\partial s_3}{\partial \rho}\right),
\end{aligned}
\end{equation}
along with the functions $s_{01}\equiv s_{01}(\varepsilon^{(1)})$, $s_{02}\equiv s_{02}(\varepsilon^{(2)})$, 
$s_i\equiv s_i(\rho,c)$  ($i=1,2,3$), $\phi_1\equiv\phi_1(\rho c)=\phi_1(\rho^{(1)})$ and 
$\phi_2\equiv\phi_2(\rho(1-c))=\phi_2(\rho^{(2)})$.

The principle of maximum entropy at 
equilibrium is satisfied if (\ref{specificentropy}) is such that  
the quadratic part in the gradients
is negative semidefinite, whence
\begin{equation}
\begin{aligned}
&\frac{\partial^3 s_1}{\partial\rho^3}\le 0,\qquad
\rho\frac{\partial^2 s_3}{\partial \rho^2}+2\frac{\partial s_3}{\partial \rho}\le 0, \\
&\frac{\partial^3 s_1}{\partial\rho^3}\left(\rho\frac{\partial^2 s_3}{\partial \rho^2}+2\frac{\partial s_3}{\partial \rho}\right)-
\frac{\rho}{4}\left(\frac{\partial^3 s_2}{\partial\rho^3}\right)^2\ge 0.
\end{aligned}
\end{equation}

It is worth noticing that the product of the mixture density times the equilibrium part of the specific entropy is the sum 
of two contributions corresponding to the two fluids; each contribution, in fact, depends on the partial mass density and 
internal energy of the corresponding fluid. On the contrary, such a consideration is not applicable to the non-local part of the specific entropy, where the contributions of the two fluids are strictly tangled. In situations close to the equilibrium, we can introduce 
the partial absolute temperatures of the two fluids, say
\begin{equation}\label{partialtheta}
\frac{1}{\theta^{(1)}}=c \frac{d s_{01}}{d\varepsilon^{(1)}},\qquad
\frac{1}{\theta^{(2)}}=(1-c) \frac{d s_{02}}{d\varepsilon^{(2)}},
\end{equation}
and the absolute temperature $\theta$ of the whole mixture can be defined as
\begin{equation}
\theta=c \theta^{(1)}+(1-c)\theta^{(2)}=\left(\frac{d s_{01}}{d\varepsilon^{(1)}}\right)^{-1}+\left(\frac{d s_{02}}{d\varepsilon^{(2)}}\right)^{-1}.
\end{equation}
Furthermore, by considering relations (\ref{partialtheta}), 
if the second order derivatives of $s_{01}(\varepsilon^{(1)})$ and
$s_{02}(\varepsilon^{(2)})$ are non-vanishing, by using the implicit function theorem, the internal energies $\varepsilon^{(1)}$ and $\varepsilon^{(2)}$ can be expressed as functions of  
$\hat{\theta}^{(1)}=c\theta^{(1)}$, and $\hat{\theta}^{(2)}=(1-c)\theta^{(2)}$, respectively.
Consequently, the partial heat fluxes take the form
\begin{equation}
\begin{aligned}
\mathbf{q}^{(a)}&=q^{(a)}_1c\frac{d\varepsilon^{(1)}}{d\hat{\theta}^{(1)}}\nabla\theta^{(1)}+q^{(a)}_2(1-c)\frac{d\varepsilon^{(2)}}{d\hat{\theta}^{(2)}}\nabla\theta^{(2)}\\
&+\left(q^{(a)}_1\theta^{(1)}\frac{d\varepsilon^{(1)}}{d\hat{\theta}^{(1)}}-q^{(a)}_2\theta^{(2)}\frac{d\varepsilon^{(2)}}{d\hat{\theta}^{(2)}}\right)\nabla c .
\end{aligned}
\end{equation}
The first  two terms describe Fourier-like effects, whereas the third one  provides  a contribution to the partial heat fluxes due to the gradient of concentration.
In the particular case where 
\begin{equation}
q_1^{(a)}\theta^{(1)}\frac{d \varepsilon^{(1)}}{d\hat{\theta}^{(1)}}
=q_2^{(a)}\theta^{(2)}\frac{d \varepsilon^{(2)}}{d\hat{\theta}^{(2)}},\qquad a=1,2,
\end{equation}
the constitutive laws for the partial heat fluxes reduce to
\begin{equation}
\mathbf{q}^{(a)}=q_1^{(a)}\frac{d\varepsilon^{(1)}}{d\hat{\theta}^{(1)}}\left(c\nabla\theta^{(1)}+(1-c)\frac{\theta^{(1)}}{\theta^{(2)}}\nabla\theta^{(2)}\right),
\end{equation}
whereas, if the more restrictive assumption
\[
q^{(a)}_1\frac{d\varepsilon^{(1)}}{d\hat{\theta}^{(1)}}=q^{(a)}_2\frac{d\varepsilon^{(2)}}{d\hat{\theta}^{(2)}}=q^{(a)}, \qquad a=1,2,
\]
is satisfied, then 
\begin{equation}
\mathbf{q}^{(a)}=q^{(a)}\nabla\left(c\theta^{(1)}+(1-c)\theta^{(2)}\right)=q^{(a)}\nabla\theta.
\end{equation}
\emph{i.e.}, we recognize the classical Fourier law of heat conduction.

Moreover, the following expressions for the material functions entering the two partial Cauchy stress tensors are determined:
\begin{align*}
\tau^{(1)}_0&=\theta^{(1)} c\left(\rho c\phi_1^\prime-\phi_1\right),\qquad \tau^{(2)}_0=\theta^{(2)}(1-c)\left(\rho(1-c)\phi_2^\prime-\phi_2\right),\allowdisplaybreaks\\ 
\tau^{(a)}_1&=\theta^{(a)}(c^{(a)})^2\left(\overline{c}^{(a)}\left(\frac{\partial^3 s_1}{\partial \rho^2 \partial c} -\rho \frac{\partial^4 s_1}{\partial 
\rho^3\partial c}
+\rho^2 \frac{\partial}{\partial\rho}\left(\rho^2\frac{\partial^3 s_2}{\partial \rho^3}-\frac{\partial s_2}{\partial \rho}\right)\right)\right.\\
&\left.-\frac{\partial}{\partial\rho}\left(\rho^2\frac{\partial^3 s_1}{\partial \rho^3}\right)\right), \allowdisplaybreaks\\
\tau^{(a)}_2&=2\theta^{(a)}\rho (c^{(a)})^2\left(-\frac{\partial}{\partial\rho}\left(\rho \frac{\partial^3 s_1 }{\partial \rho^2\partial c}\right)
+\frac{\partial^2 s_2 }{\partial\rho^2}
+\overline{c}^{(a)}\frac{\partial^3}{\partial\rho^3}\left(\rho^2 s_3\right)\right), \allowdisplaybreaks\\
\tau^{(a)}_3&=\theta^{(a)} \rho^2 c^{(a)}\left(\frac{\partial^2 s_2}{\partial \rho^2}-\frac{\partial^3s_1 }{\partial\rho^2\partial c}-\rho c^{(a)}\frac{\partial^4s_2}
{\partial\rho^3\partial c}+\rho^2c^{(a)}\frac{\partial^3s_3}{\partial\rho^3}\right. +\rho c^{(1)}c^{(2)}\frac{\partial^3s_3}{\partial\rho^2\partial c}\\
&+\left.\rho(5c^{(a)}+\overline{c}^{(a)})\frac{\partial^2s_3}{\partial \rho^2}+2c^{(1)}c^{(2)}\frac{\partial^2 s_3}{\partial\rho\partial c}+(4c^{(a)}+2\overline{c}^{(a)})\frac{\partial s_3}{\partial\rho}\right), \allowdisplaybreaks\\
\tau^{(a)}_4&=2\theta^{(a)}\rho (c^{(a)})^3\left(-\frac{\partial^3 s_1}{\partial\rho^2\partial c}+\frac{\partial^2 s_2}{\partial\rho^2}+\rho\overline{c}^{(a)}\frac{\partial^2 s_3}{\partial\rho^2}+2\overline{c}^{(a)}\frac{\partial s_3}{\partial\rho}\right),\allowdisplaybreaks \\
\tau^{(a)}_5&=2\theta^{(a)}\rho^2 (c^{(a)})^2\left(-\frac{\partial^3 s_1}{\partial\rho^2\partial c}+\frac{\partial^2 s_2}{\partial\rho^2}+\rho\overline{c}^{(a)}\frac{\partial^2 s_3}{\partial\rho^2}+2\overline{c}^{(a)}\frac{\partial s_3}{\partial\rho}\right), \allowdisplaybreaks\\
\tau^{(a)}_7&=\theta^{(a)}\rho (c^{(a)})^2\left(2\frac{\partial^3 s_1}{\partial\rho^3}-\overline{c}^{(a)}\frac{\partial^3 s_2}{\partial\rho^3}\right),\allowdisplaybreaks \\
\tau^{(a)}_8&=4\theta^{(a)}\rho (c^{(a)})^2\left(\frac{\partial^3 s_1}{\partial\rho^2\partial c}-\frac{\partial^2 s_2}{\partial\rho^2}-\overline{c}^{(a)}\left(\rho\frac{\partial^2s_3}{\partial \rho^2}+2\frac{\partial s_3}{\partial\rho}\right)\right), \allowdisplaybreaks\\
\tau^{(a)}_9 &=2\theta^{(a)}\rho^2 c^{(a)}\left(\frac{\partial^3 s_1}{\partial\rho^2\partial c}-\frac{\partial^2 s_2}{\partial\rho^2}-\overline{c}^{(a)}\left(\rho\frac{\partial^2s_3}{\partial \rho^2}+2\frac{\partial s_3}{\partial\rho}\right)\right), \allowdisplaybreaks\\
\tau^{(a)}_{10} &=\tau^{(a)}_{11}=0,
\end{align*}
where the prime ${}^\prime$ denotes differentiation with respect to the argument, $c^{(1)}=\overline{c}^{(2)}=c$, $c^{(2)}=\overline{c}^{(1)}=c-1$, and $a=1,2$. 

The scalar functions $s_i$ $(i=1,2,3)$, involved in the expression of the entropy density and in the partial Cauchy stress tensors, have to satisfy the following additional constraints:
\begin{eqnarray}
&&2\frac{\partial^3 s_1}{\partial\rho^2\partial c}-\rho\frac{\partial^3 s_2}{\partial\rho^3}-2\frac{\partial^2 s_2}{\partial\rho^2}=0, \label{constr1}
\end{eqnarray}
\begin{eqnarray}
&&4c^3\left(\rho\frac{\partial^2 s_1}{\partial\rho^2}-\frac{\partial s_1}{\partial\rho}\right)+2c^3\left(c(1-c)\left(\rho \frac{\partial^3 s_2}{\partial\rho^2\partial c}-\frac{\partial^2 s_2}{\partial\rho\partial c}\right)\right.\nonumber\\
&&\qquad\left.+(1-2c)\left(\rho\frac{\partial^2 s_2}{\partial\rho^2}-\frac{\partial s_2}{\partial\rho}\right)\right)-4c^4(1-c)\frac{\partial}{\partial\rho}\left(\rho^2\frac{\partial s_3}{\partial\rho}+\rho  s_3\right)\nonumber\\
&&\qquad-c^3\psi_2^\prime-\psi_1=0, \label{constr2}
\end{eqnarray}
\begin{eqnarray}
&&  2c^3\left(\rho^2\frac{\partial^3 s_1}{\partial\rho^3}+\rho(1-2c)\frac{\partial^3 s_1}{\partial\rho^2\partial c}-c(1-c)\frac{\partial^3 s_1}{\partial\rho \partial c^2}-2(1-2c)\frac{\partial^2 s_1}{\partial\rho\partial c}\right) \nonumber \\
&& \qquad+2c^3\left(c(1-c)\frac{\partial^2 s_2}{\partial\rho\partial c}+(1-2c)\frac{\partial s_2}{\partial\rho}\right)+c^3\psi_2^\prime+\psi_1=0, \label{constr3}
\end{eqnarray}
\begin{eqnarray}
&& 2c^3\left(\rho^2\frac{\partial^3 s_1}{\partial\rho^3}-\rho(1-2c)\frac{\partial^3 s_1}{\partial\rho^2\partial c}+c(1-c)\frac{\partial^3 s_1}{\partial\rho \partial c^2}+2(1-2c)\frac{\partial^2 s_1}{\partial\rho \partial c}\right)\nonumber  \\
&& \qquad+2c^3\left(\rho^2(1-2c)\frac{\partial^3 s_2}{\partial\rho^3}-c(1-c)\frac{\partial^2 s_2}{\partial\rho \partial c}-(1-2c)\frac{\partial s_2}{\partial\rho}\right) \nonumber \\
&& \qquad -4\rho c^4(1-c)\left(\rho\frac{\partial^2 s_3}{\partial\rho^2}+2\frac{\partial s_3}{\partial\rho}\right)-c^3\psi_2^\prime-\psi_1=0, \label{constr4}
\end{eqnarray}
\begin{eqnarray}
&&c^5(1-c)^2\left(c(1-c)\frac{\partial^{4}s_{1}}{\partial c^4}-\rho(1-2c)\frac{\partial^{4}s_{1}}{\partial\rho\partial c^3}-\rho^2 \frac{\partial^{4}s_{1}}{\partial\rho^2\partial c^2}+5 (1-2c)\frac{\partial^{3}s_{1}}{\partial c^3}\right.\nonumber\\
&& \qquad+4\rho \frac{\partial^{3}s_{1}}{\partial\rho\partial c^2}-12 \frac{\partial^{2}s_{1}}{\partial c^2}+\rho^3 \frac{\partial^{4}s_{2}}{\partial\rho^3\partial c}+\rho^2 (1-2c)\frac{\partial^{4}s_{2}}{\partial\rho^2\partial c^2}-\rho c(1-c)\frac{\partial^{4}s_{2}}{\partial\rho\partial c^3}\nonumber\\
&& \qquad-3\rho^2 \frac{\partial^{3}s_{2}}{\partial\rho^2\partial c}-5\rho (1-2c)\frac{\partial^{3}s_{2}}{\partial\rho\partial c^2}+c(1-c)\frac{\partial^{3}s_{2}}{\partial c^3}+12\rho\frac{\partial^{2}s_{2}}{\partial\rho\partial c} \nonumber\\
&& \qquad+5(1-2c)\frac{\partial^{2}s_{2}}{\partial c^2}-12\frac{\partial s_{2}}{\partial c}-\rho^4\frac{\partial^3 s_{3}}{\partial \rho^3}-\rho^3(1-2c)\frac{\partial^3 s_{3}}{\partial \rho^2\partial c} \nonumber\\
&& \qquad+\rho^2 c(1-c)\frac{\partial^3 s_{3}}{\partial \rho\partial c^2}-2\rho^3\frac{\partial^2 s_{3}}{\partial \rho^2}+\rho^2 (1-2c)\frac{\partial^2 s_{3}}{\partial \rho\partial c}+2\rho c(1-c)\frac{\partial^2 s_{3}}{\partial c^2} \nonumber \\
&& \qquad-\left.2\rho^2 \frac{\partial s_{3}}{\partial \rho}+4\rho (1-2c)\frac{\partial s_{3}}{\partial c}-4\rho s_3\right)=0, 
\label{constr5}
\end{eqnarray}
\begin{eqnarray}
&&  4c^3\left(c^2(1-c^2)\frac{\partial^3 s_1}{\partial c^3}-2\rho^2(1-2c)\frac{\partial^2 s_1}{\partial\rho^2}+\rho(2-7c(1-c))\frac{\partial^2 s_1}{\partial\rho\partial c}\right. \nonumber \\
&& \qquad +\left.4c(1-c)(1-2c)\frac{\partial^2 s_1}{\partial c^2}+2\rho(1-2c)\frac{\partial s_1}{\partial\rho}-4c(1-c)\frac{\partial s_1}{\partial c}\right) \nonumber \\
&&  \qquad -2c^3\left(\rho c^2(1-c)^2\frac{\partial^3 s_2}{\partial\rho\partial c^2}+2\rho^2(2-7c(1-c))\frac{\partial^2 s_2}{\partial\rho^2}\right.\nonumber\\
&&\qquad\left.+6\rho c(1-c)\left((1-2c)\frac{\partial^2 s_2}{\partial\rho\partial c} -\frac{\partial s_2}{\partial\rho}\right)-4c(1-c)\left((1-2c)\frac{\partial s_2}{\partial c}-s_2\right)\right) \nonumber \\
&& \qquad +4\rho c^4(1-c)\left(3\rho(1-2c)\frac{\partial s_3}{\partial\rho}+c(1-c)\frac{\partial s_3}{\partial c}+3(1-2c)s_3\right) \nonumber \\
&& \qquad -(1-c)\left(\rho(c \psi_1^\prime-3\psi_1)+c^4(\rho \psi_2^{\prime\prime}+\psi_3^{\prime\prime})\right)=0, \label{constr6}
\end{eqnarray}
\begin{eqnarray}
&&  2c^3\left(c^2(1-c)^2\frac{\partial^3 s_1}{\partial c^3}-2\rho^2(1-2c)\frac{\partial^2 s_1}{\partial\rho^2}+2\rho(1-2c)^2\frac{\partial^2 s_1}{\partial\rho\partial c}\right. \nonumber \\
&& \qquad+\left.4c(1-c)(1-2c)\frac{\partial^2 s_1}{\partial c^2}+2\rho(1-2c)\frac{\partial s_1}{\partial\rho}-4c(1-c)\frac{\partial s_1}{\partial c}\right) \nonumber \\
&& \qquad-2c^3\left(2\rho(1-2c)\left(\rho(1-2c)\frac{\partial^2 s_2}{\partial\rho^2}+ c(1-c)\frac{\partial^2 s_2}{\partial\rho\partial c}\right)\right. \nonumber \\
&& \qquad \left.+c(1-c)\left(c(1-c)\frac{\partial^2 s_2}{\partial c^2}-2\rho\frac{\partial s_2}{\partial\rho}\right)\right)\nonumber\\
&&\qquad-2\rho c^4(1-c)\left(\rho c(1-c)\frac{\partial^2 s_3}{\partial\rho\partial c}-2(1-2c)\left(\rho\frac{\partial s_3}{\partial\rho}+s_3\right)\right) \nonumber \\
&& \qquad-(1-c)\left(\rho(c \psi_1^\prime-3\psi_1)+c^4(\rho \psi_2^{\prime\prime}+\psi_3^{\prime\prime})\right)=0,\label{constr7}
\end{eqnarray}
where $\psi_i\equiv\psi_i(c)$ ($i=1,2,3$) are arbitrary functions of the indicated argument.

Furthermore, by requiring in the entropy inequality the condition $\mathbf{C}=\mathbf{0}$, \emph{i.e.}, vanishing the coefficients of linear terms in the higher
derivatives, we obtain the following representation of the entropy flux: 
\begin{equation}\label{entropyflux}
\begin{aligned}
\mathbf{J}&=\frac{\mathbf{q}^{(1)}}{c\theta^{(1)}}+\frac{\mathbf{q}^{(2)}}{(1-c)\theta^{(2)}}
+\left(\rho c(s_{01}-s_{02})+\phi_1 -\frac{c}{1-c}\phi_2\right)\mathbf{w} \\
&+\left(2\rho^2\hat{s}_1(\nabla\cdot\mathbf{v})+\rho c\hat{s}_2(\nabla\cdot\mathbf{w})\right)\nabla\rho \\
&+\left(\rho^2\hat{s}_2(\nabla\cdot\mathbf{v})+2\rho c\hat{s}_3(\nabla\cdot\mathbf{w})\right)\nabla c \\
&+\left(\frac{c}{2}\hat{s}_2|\nabla\rho|^2+2c\hat{s}_3(\nabla\rho\cdot\nabla c)+\rho\frac{\rho\hat{s}_2+2(1-2c)\hat{s}_3}{2(1-c)}|\nabla c|^2\right)\mathbf{w} \\
&+(\rho\hat{s}_2-2c\hat{s}_3)(\nabla c\cdot\mathbf{w})\nabla\rho-\rho\frac{\rho\hat{s}_2-2c\hat{s}_3}{1-c}(\nabla c\cdot\mathbf{w})\nabla c,
\end{aligned}
\end{equation}
where relations (\ref{solentropy}) and (\ref{constr1})--(\ref{constr7}) have been used.

In (\ref{entropyflux}) we recognize the contribution of the classical terms $\displaystyle \frac{\mathbf{q^{(1)}}}{c\theta^{(1)}}$ and $\displaystyle \frac{\mathbf{q^{(2)}}}{(1-c)\theta^{(2)}}$, and an entropy extra-flux \cite{Mul} depending on the diffusion velocity, the divergence of the barycentric and diffusion velocities, together with  the first order gradients of the mass density of the mixture and the concentration of the first constituent. Therefore, the extra-flux, as one expects, includes the effects due to the interaction of the two fluids, and the diffusion velocity plays an important role. However, even if diffusive effects are neglected (\emph{i.e}, $\mathbf{w}=\mathbf{0}$), the entropy extra-flux does not vanish, and is similar to the one derived in Ref.~\cite{GP-2021} for a single Korteweg fluid.

The matrix $\mathbf{B}$, with the above results, identically vanishes, and
the entropy inequality reduces to 
\begin{equation}\label{finalineq}
\begin{aligned}
&\mathbf{q}^{(1)}\cdot \nabla\left(\frac{d s_{01}}{d\varepsilon^{(1)}}\right)+\mathbf{q}^{(2)}\cdot \nabla\left(\frac{d s_{02}}{d\varepsilon^{(2)}}\right) \\
&+\left(\tau^{(1)}_6 (\nabla\cdot(\mathbf{v}+\mathbf{w}))^2+\tau^{(1)}_{12} \left(\nabla(\mathbf{v}+\mathbf{w})\cdot\nabla(\mathbf{v}+\mathbf{w})\right)\right)\frac{d s_{01}}{d\varepsilon^{(1)}} \\
&+\left(\tau^{(2)}_6\left(\nabla\cdot\left(\mathbf{v}-\frac{c}{1-c}\mathbf{w}\right)\right)^2\right. \\
&+\left.\tau^{(2)}_{12} \left(\nabla\left(\mathbf{v}-\frac{c}{1-c}\mathbf{w}\right)\cdot\nabla\left(\mathbf{v}-\frac{c}{1-c}\mathbf{w}\right)\right)\right)\frac{d s_{02}}{d\varepsilon^{(2)}}\ge 0.
\end{aligned}
\end{equation}
The residual entropy inequality (\ref{finalineq}) turns out to be a homogeneous quadratic polynomial in some gradients entering the state space, whose coefficients depend at most on the field variables; such a relation is satisfied for all the thermodynamical processes if and only if the following inequalities hold true:
\begin{equation}\label{ineqfinal}
\begin{aligned}
&q^{(1)}_1 s_{01}^{\prime\prime}\geq 0,\qquad q^{(2)}_2s_{02}^{\prime\prime}\geq 0, \\
&\tau_6^{(1)}s^{\prime}_{01}\geq 0,\qquad \tau_{12}^{(1)}s^{\prime}_{01}\ge 0, \qquad
\tau_6^{(2)}s^{\prime}_{02}\geq 0,\qquad \tau_{12}^{(2)}s^{\prime}_{02}\ge 0, \\
&4q^{(1)}_1q^{(2)}_2s_{01}^{\prime\prime}s_{02}^{\prime\prime}-\left(q^{(1)}_2s_{01}^{\prime\prime}+q^{(2)}
_1s_{02}^{\prime\prime}\right)^2\ge 0.
\end{aligned}
\end{equation}
Since
\begin{equation}
\begin{aligned}
&s_{01}^{\prime}=\frac{1}{c\theta^{(1)}}>0,\qquad
s_{02}^{\prime}=\frac{1}{(1-c)\theta^{(2)}}>0, \\
&s_{01}^{\prime\prime}=-\frac{1}{c(\theta^{(1)})^2}\frac{\partial\theta^{(1)}}{\partial\varepsilon^{(1)}}<0,\qquad
s_{02}^{\prime\prime}=-\frac{1}{(1-c)(\theta^{(2)})^2}\frac{\partial\theta^{(2)}}{\partial\varepsilon^{(2)}}<0,
\end{aligned}
\end{equation}
the inequalities (\ref{ineqfinal}) imply also
\begin{equation*}
q^{(1)}_1\le 0,\qquad q^{(2)}_2\le 0, \qquad
\tau_6^{(1)}\ge 0,\qquad\tau_{12}^{(1)}\ge 0, \qquad
\tau_6^{(2)}\ge 0,\qquad\tau_{12}^{(2)}\ge 0,
\end{equation*}
that are physically admissible.

Thus, the results above detailed complete the exploitation of entropy inequality 
for the three-dimensional case of a binary mixture of third grade viscous Korteweg fluids with two velocities and two temperatures.

As a final remark, we note that all the constitutive equations above determined must be such that the differential constraints (\ref{constr1})--(\ref{constr7}), that cannot be solved in general,  are satisfied. Nevertheless, we are able to prove that they are compatible by exhibiting a particular solution
characterizing the scalar functions $s_i$ ($i=1,2,3$), $\psi_1$ and $\psi_3$:
\begin{equation}
\begin{aligned}
&s_1=\rho\left(\kappa_1(1-\log(\rho))+\rho \chi_1+\chi_2\right), \\
&s_2=\rho\left(\rho\chi_1^\prime+\chi_3-\kappa_2\frac{\log(\rho)}{\rho c(1-c)}\right), \\
&s_3=\frac{1}{2c(1-c)}\left(-c(1-c)\chi_2^{\prime\prime}-2(1-2c)\chi_2^\prime-2\chi_2+\kappa_1(2\log(\rho)-5)\right. \\
&\quad-2\kappa_2\frac{16c^2(1-c)^2\arctanh(1-2c)-(1-2c)(\log(\rho)-4c(1-c)-1)}{\rho c(1-c)}, \\
&\psi_1=c^3\left(-2\kappa_1+2c(1-c)(\chi_2^{\prime\prime}-\chi_3^{\prime})+2(1-2c)(2\chi_2^\prime-\chi_3)-\psi_2^{\prime}\right), \\
&\psi_3=-4\kappa_2\left(2c\,\arctanh(1-2c)+\log(c)\right)+\kappa_3 c+\kappa_4,
\end{aligned}
\end{equation}
$\chi_i$ ($i=1,2,3$) being arbitrary functions depending on $c$, and $\kappa_i$ ($i=1,\ldots,4$) arbitrary constants; in such a way, conditions (\ref{constr1})--(\ref{constr7}) are identically satisfied.

Because of  these positions, the specific entropy can be written in the form
\begin{equation}
\begin{aligned}
s&=\frac{1}{\rho}\left(\rho c s_{01}+\rho(1-c)s_{02}+\phi_1+\phi_2\right) \\
&+\frac{\kappa_1}{\rho^2}|\nabla\rho|^2-2\frac{\kappa_2}{\rho^2c(1-c)}\nabla\rho\cdot\nabla c+\frac{\kappa_1\rho c(1-c)-\kappa_2(1-2c)}{\rho c^2(1-c)^2}|\nabla c|^2,
\end{aligned}
\end{equation}
and the principle of maximum entropy at equilibrium holds true in the following two cases:
\begin{itemize}
\item $\kappa_2\le 0$ and $\displaystyle\kappa_1\le \frac{\kappa_2}{\rho c}$;
\item $\kappa_2>0$ and $\displaystyle\kappa_1\le\frac{\kappa_2}{\rho(c-1)}$.
\end{itemize}
Furthermore, the material functions entering the partial Cauchy stress tensors simplify to
\begin{equation}
\label{simpletau}
\begin{aligned}
\tau^{(1)}_0&=\theta^{(1)} c\left(\rho c\phi_1^\prime-\phi_1\right),\qquad &&\tau^{(2)}_0=\theta^{(2)}(1-c)\left(\rho(1-c)\phi_2^\prime-\phi_2\right), \\
\tau^{(1)}_1&=-\kappa_2\theta^{(1)}\frac{c}{\rho^2},\qquad&&\tau^{(2)}_1=\kappa_2\theta^{(2)}\frac{1-c}{\rho^2}, \\
\tau^{(1)}_2&=2\theta^{(1)}\frac{\kappa_2-2\kappa_1\rho c}{\rho},\qquad
&&\tau^{(2)}_2=2\theta^{(2)}\frac{\kappa_2+2\kappa_1\rho(1-c)}{\rho}, \\
\tau^{(1)}_3&=-\kappa_2\frac{\theta^{(1)}}{c},\qquad &&\tau^{(2)}_3=\kappa_2\frac{\theta^{(2)}}{1-c}, \\
\tau^{(1)}_4&=2\theta^{(1)}\frac{c}{\rho}(\kappa_2-\kappa_1\rho c),\qquad
&&\tau^{(2)}_4=-2\theta^{(2)}\frac{1-c}{\rho}(\kappa_2+\kappa_1\rho(1-c)), \\
\tau^{(1)}_5&=2\theta^{(1)}(\kappa_2-\kappa_1\rho c),\qquad
&&\tau^{(2)}_5=2\theta^{(2)}(\kappa_2+\kappa_1\rho(1-c)), \\
\tau^{(1)}_7&=-2\theta^{(1)}\frac{c}{\rho^2}(\kappa_2-\kappa_1\rho c),\qquad
&&\tau^{(2)}_7=2\theta^{(2)}\frac{1-c}{\rho^2}(\kappa_2+\kappa_1\rho(1-c)), \\
\tau^{(1)}_8&=-4\theta^{(1)}\frac{\kappa_2-\kappa_1\rho c}{\rho},\qquad
&&\tau^{(2)}_8=-4\theta^{(2)}\frac{\kappa_2+\kappa_1\rho(1-c)}{\rho}, \\
\tau^{(1)}_9&=-2\theta^{(1)}\frac{\kappa_2-\kappa_1\rho c}{c},\qquad
&&\tau^{(2)}_9=2\theta^{(2)}\frac{\kappa_2+\kappa_1\rho(1-c)}{1-c},
\end{aligned}
\end{equation}
and the entropy flux $\mathbf{J}$, given in (\ref{entropyflux}), reduces to:
\begin{equation}
\begin{aligned}
\mathbf{J}&=\frac{\mathbf{q}^{(1)}}{c\theta^{(1)}}+\frac{\mathbf{q}^{(2)}}{(1-c)\theta^{(2)}}
+\left(\rho c(s_{01}-s_{02})+\phi_1 -\frac{c}{1-c}\phi_2\right)\mathbf{w} \\
&+\left(2\kappa_1(\nabla\cdot\mathbf{v})-2\frac{\kappa_2}{\rho(1-c)}(\nabla\cdot\mathbf{w})\right)\nabla\rho \\
&+\left(-2\frac{\kappa_2}{c(1-c)}(\nabla\cdot\mathbf{v})+2\frac{\kappa_1\rho c(1-c)-\kappa_2(1-2c)}{c(1-c)^2}(\nabla\cdot\mathbf{w})\right)\nabla c \\
&+\left(-\frac{\kappa_2}{\rho^2(1-c)}|\nabla\rho|^2+2\frac{\kappa_1\rho c(1-c)-\kappa_2(1-2c)}{\rho c(1-c)^2}(\nabla\rho\cdot\nabla c)\right. \\
&+\left.\frac{\kappa_1\rho c(1-c(3-2c))-\kappa_2(1-3c(1-c))}{c^2(1-c)^3}|\nabla c|^2\right)\mathbf{w} \\
&-2\frac{\kappa_1\rho(1-c)+\kappa_2}{\rho(1-c)^2}(\nabla c\cdot\mathbf{w})\nabla\rho+2\frac{\kappa_1\rho(1-c)+\kappa_2}{(1-c)^3}(\nabla c\cdot\mathbf{w})\nabla c.
\end{aligned}
\end{equation}

\section{Equilibrium conditions}
\label{sec:serrin3D}
By using the expression of the constitutive laws provided in the previous section, let us consider the problem of determining the phase boundaries at the equilibrium for the mixture on a purely mechanical framework. At equilibrium, the partial temperatures are constant and equal, and $\mathbf{v}=\mathbf{w}=\mathbf{0}$.
Before analyzing this problem, let us recall a general result established for the phase boundaries at equilibrium of a Korteweg fluid.
The search for equilibrium configurations of a Korteweg fluid whose Cauchy stress tensor is given by (\ref{kort}) requires to solve the condition
\begin{equation}
\label{equil-single}
\nabla\cdot\left(\left(-p+\alpha_1|\nabla\rho|^2+\alpha_2\Delta\rho\right)\mathbf{I}+\alpha_3\nabla\rho\otimes\nabla\rho+\alpha_4\nabla\nabla\rho\right)=\mathbf{0},
\end{equation}
where the pressure $p$ and the material functions $\alpha_i$ ($i=1,\ldots,4$) now depend only on the mass density $\rho$.
Condition (\ref{equil-single}) has three independent components while liquid-vapor phase equilibria are 
determined by just one physical variable, namely the mass density,  \emph{i.e.}, the equilibrium system (\ref{equil-single}) 
is overdetermined.
A remarkable result obtained in 1983 by Serrin \cite{Serrin} proved that, unless rather special conditions are satisfied, the only geometric phase boundaries which are consistent with relation (\ref{equil-single}) are either spherical, cylindrical, or 
planar. Using a general theorem proved in Ref.~\cite{Pucci}, Serrin was able to prove that the constitutive 
equation (\ref{kort}) must be such that the coefficients therein involved must obey the condition
\begin{equation}
\label{condkort}
\begin{aligned}
&(\alpha_1+\alpha_3) \left(\frac{\partial\alpha_4}{\partial\rho}-\alpha_3\right)\\
&\qquad +\frac{1}{2}\left(\left(\frac{\partial\alpha_4}{\partial\rho}-\alpha_3\right)^2-(\alpha_2+\alpha_4)\frac{\partial}{\partial \rho}
\left(\frac{\partial\alpha_4}{\partial\rho}-\alpha_3\right)\right)=0,
\end{aligned}
\end{equation}
in order to avoid \cite{Pucci} that the solution of (\ref{equil-single}) possesses solutions described only by
level surfaces with constant mean and Gaussian curvature \cite{MOV,GO-2017,GO-2019}, 
which are either (pieces of) concentric spheres, or concentric circular cylinders, or parallel planes.
From the physical viewpoint, this  result reflects the experimental evidence that several (but not all) phase 
boundaries  have constant mean curvature.
On the other hand, the question can be raised whether equation (\ref{condkort}) is  physically necessary, since 
without it the theory allows very few equilibrium configurations. It is worth observing that, although rather unusual, 
this fact  surely does not necessarily lead to the condition (\ref{condkort}).  Moreover, as it turns out by the results of 
Section~\ref{sec:liu},  the second law of thermodynamics, in the form of the generalized Clausius-Duhem inequality 
(\ref{entropyconstrained}), does not require (\ref{condkort}). 
Under these circumstances, Serrin argued that his result only offers significant  reasons to accept the restriction (\ref{condkort}) for any physically realistic Korteweg fluid \cite{Serrin}, although it is not a necessary condition for the equilibrium. We remark that in Ref.~\cite{GP-2021}, where a complete thermodynamical analysis of viscous Korteweg fluids has been done, the constitutive functions therein determined can be chosen in such a way the Serrin condition is fulfilled.

Let us now focus on mechanical equilibrium configurations for a mixture of two Korteweg fluids, \emph{i.e.}, we have to investigate the relations:
\begin{equation}
\label{equil-mixture}
\begin{aligned}
&\nabla\cdot\left(
\left(\tau^{(a)}_0+\tau^{(a)}_1 |\nabla\rho|^2+\tau^{(a)}_2 \nabla\rho\cdot\nabla c+\tau^{(a)}_3 |\nabla c|^2+\tau^{(a)}_4 \Delta\rho+\tau^{(a)}_5 \Delta c\right)\mathbf{I}\right. \\
&\qquad+\tau^{(a)}_7\nabla\rho\otimes\nabla\rho+\tau^{(a)}_8\sym\left(\nabla\rho\otimes\nabla c\right)
+\tau^{(a)}_9\nabla c\otimes\nabla c \\
&\qquad\left.+\tau^{(a)}_{10}\nabla\nabla\rho+\tau^{(a)}_{11}\nabla\nabla c\right)=\mathbf{0},
\qquad a=1,2;
\end{aligned}
\end{equation}
notice that we have  an overdetermined system with six scalar differential equations for the two unknowns $\rho$ and $c$.

Taking into account the expressions of the material functions $\tau^{(a)}_i$, along with the differential constraints (\ref{constr1})--(\ref{constr7}), we notice that the conditions
(\ref{equil-mixture}) can be written as
\begin{equation}\label{equil-fluids}
\begin{aligned}
&\nabla\cdot\left(\left(-p^{(a)}+\alpha^{(a)}_1|\nabla\rho^{(a)}|^2+\alpha^{(a)}_2\Delta\rho^{(a)}\right)\mathbf{I}\right.\\ 
&\qquad\left.+\alpha^{(a)}_3\nabla\rho^{(a)}\otimes\nabla\rho^{(a)}+\alpha^{(a)}_4\nabla\nabla\rho^{(a)}\right)=\mathbf{0}, \qquad a=1,2,
\end{aligned}
\end{equation}
where 
\begin{equation}
\label{ouralpha}
\begin{aligned}
& p^{(1)}=-\tau^{(1)}_0, \quad &&p^{(2)}=-\tau^{(2)}_0,\\
&\alpha^{(1)}_1=\frac{\tau^{(1)}_1}{c^2}=\frac{\tau^{(1)}_3}{\rho^2}, \quad  &&\alpha^{(2)}_1=\frac{\tau^{(2)}_1}{(1-c)^2}=\frac{\tau^{(2)}_3}{\rho^2}, \\
&\alpha^{(1)}_2=\frac{\tau^{(1)}_4}{c}=\frac{\tau^{(1)}_5}{\rho}=\frac{\tau^{(1)}_2}{2}-\alpha^{(1)}_1\rho c, \quad &&\alpha^{(2)}_2=\frac{\tau^{(2)}_4}{1-c}=-\frac{\tau^{(2)}_5}{\rho}=-\frac{\tau^{(2)}_2}{2}-\alpha^{(2)}_1\rho(1-c), \\
&\alpha^{(1)}_3=\frac{\tau^{(1)}_7}{c^2}=\frac{\tau^{(1)}_8}{2\rho c}=\frac{\tau^{(1)}_9}{\rho^2},\quad &&\alpha^{(2)}_3=\frac{\tau^{(2)}_7}{(1-c)^2}=-\frac{\tau^{(2)}_8}{2\rho(1-c)}=\frac{\tau^{(2)}_9}{\rho^2},\\
&\alpha^{(1)}_4=\tau^{(1)}_{10}=\tau^{(1)}_{11},\quad &&\alpha^{(2)}_4=\tau^{(2)}_{10}=\tau^{(2)}_{11}.
\end{aligned}
\end{equation} 

Therefore, the equilibrium conditions are somehow decoupled, even if the coefficients therein involved depend on $\rho$ and $c$. 
In such a way, the Serrin conditions corresponding to the equations (\ref{equil-fluids}) read
\begin{equation}
\label{serrin-mixture}
\begin{aligned}
 &\left(\alpha^{(a)}_1+\alpha^{(a)}_3\right) \left(\frac{\partial\alpha^{(a)}_4}{\partial\rho^{(a)}}-\alpha^{(a)}_3\right)\\
 &+\frac{1}{2}\left(\left(\frac{\partial\alpha^{(a)}_4}{\partial\rho^{(a)}}-\alpha^{(a)}_3\right)^2-\left(\alpha^{(a)}_2+\alpha^{(a)}_4\right)\frac{\partial}{\partial \rho^{(a)}}
\left(\frac{\partial\alpha^{(a)}_4}{\partial\rho^{(a)}}-\alpha^{(a)}_3\right)\right)=0,
\end{aligned}
\end{equation}
for $a=1,2$. 

By inserting the expressions (\ref{ouralpha}), it is easy to recognize that the conditions 
(\ref{serrin-mixture}) are identically satisfied; then, no further restrictions to the phase boundaries at equilibrium arise; this implies that the level surfaces of phase boundaries  are not constrained to be pieces of concentric spheres, or concentric circular cylinders, or parallel planes. 
The material functions entering the partial Cauchy stress tensors here determined can be used, at least in the simpler form (\ref{simpletau}), to solve  the equilibrium conditions and compare the numerical results to the experimental ones. This will be a natural development of the present paper, and we plan to pursue this goal.

\section{Conclusions}
\label{sec:final}
In this paper, the restrictions imposed by entropy principle on the constitutive equations for a binary mixtures of viscous Korteweg fluids with two temperatures and two velocities are derived and solved, so extending the results obtained in Ref.~\cite{CGOP-2020} in the one-dimensional case; it is worth of being underlined that the model here considered includes also viscosity terms in the expression of partial Cauchy stress tensors so facing a more realistic situation with respect to the case investigated in Ref.~\cite{CGOP-2020}. Because of the non-locality of the constitutive equations, we exploited the second law of thermodynamics by means of an extended Liu method. 

First,  the field equations for the two constituents, taking into account that  they have different velocities, and so different material time derivatives, are considered. The studied model  neglects the action of external body forces, namely the gravity, and  heat 
sources, as well as momentum and energy exchanges between the components. Some of these aspects will be considered in a forthcoming paper. Then, the evolution equations for the mass density of the mixture, the concentration of one constituent, the barycentric and the diffusion velocities, and the internal energies of the constituents are obtained. It has been assumed a state space containing also second order gradients of the mixture mass density and of the concentration.
Therefore, by using the extended Liu procedure, we were able to prove that the non-local constitutive equations are compatible with second law of thermodynamics. 
Even if the computation of the thermodynamical constraints is straightforward, the intermediate results involve terribly long expression that we are forced to manage by using some symbolic routines written in the Computer Algebra System Reduce \cite{Reduce}. 

Remarkably, we were able to explicitly solve the thermodynamical constraints by obtaining Cauchy stress tensors for the two fluids that are sufficiently general to 
include that proposed by Korteweg in 1901 \cite{Kor}. 
The solution presented in this paper generalizes to the three-dimensional case the results obtained in Ref.~\cite{CGOP-2020} in one space dimension; moreover, it is complete, physically sound and immediately applicable. 

As a consequence of the procedure (which is completely algorithmic), after expanding around the equilibrium the specific entropy retaining only first order terms in the gradients of the mixture mass density and concentration of one constituent, the expression of the entropy flux naturally arises. The latter contains the classical part and an extra-flux depending on the diffusion velocity, on the gradients of mixture mass density and concentration of one constituent, and on the divergence of barycentric and diffusion velocities.
It is worth of being remarked that the extra-term in the entropy flux comes as a byproduct of the application of the extended Liu procedure without neither postulating its existence from the beginning nor modifying \emph{a priori} the energy balances or the Clausius-Duhem inequality.
Finally, we explored possible additional constraints imposed by studying the equilibrium configurations of the mixture and proved that no restrictions exist on the admissible phase boundaries. 

The theoretical results derived here contain some degrees of freedom and may serve as a basis for experimental and/or numerical investigations.  In particular, work is in progress to numerically integrate the equilibrium conditions and determine the level surfaces of phase boundaries [at least in the case of Cauchy stress tensors involving the material functions given in (\ref{simpletau})], as well as to derive a reduced evolution equation for waves propagating in a mixture of Korteweg fluids. 

Also, we plan to investigate the possibility of applying the methodology here used to liquid 
Helium 2 below the so called $\lambda$-point. In fact, in this condition, Helium 2 behaves as a mixture of a fluid component and a superfluid one \cite{Khal}, where non-local effects are detectable and non-local constitutive equations are crucial \cite{AtFox}. Thus, it could be interesting to investigate if two different temperatures could play a role and  provide a deeper insight in the physics of the problem.

\section*{Acknoledgments}
Work supported by the ``Gruppo Nazionale per la Fisica Matematica'' (GNFM) of the Istituto Nazionale di Alta 
Matematica ``F. Severi''. M.G. acknowledges the support through the ``Progetto Giovani GNFM 2020''.\\
The authors gratefully thank the anonymous referees for their valuable comments leading us to improve the quality of the paper.

\end{document}